\documentclass[12pt,aps,prd,preprint, tightenlines,superscriptaddress,amsmath,amssymb,nofootinbib]{revtex4-1} 

\usepackage[letterpaper,dvips,hmargin={1in,1in},vmargin={1in,1in}]{geometry}

\RequirePackage[colorlinks=true
,urlcolor=blue
,anchorcolor=blue
,citecolor=blue
,filecolor=blue
,linkcolor=blue
,menucolor=blue
,pagecolor=blue
,linktocpage=true
,pdfproducer=medialab
,pdfa=true
]{hyperref}

\allowdisplaybreaks

\newcommand{\newc}{\newcommand*}

\long\def\begincomment#1\endcomment{%
        \begingroup\sf\baselineskip12pt#1\endgroup}

\newc{\etal}{\textrm{et al.}} 
\newc{\eg}{\textrm{e.g.}} 
\newc{\ie}{\textrm{i.e.}}
\newc{\etc}{\textrm{etc}}
\newc\vs{\textrm{vs.}}
\newc{\cl}{\rm {C.L.}}
\newc{\ev}{\ensuremath{\,\mathrm{eV}}}
\newc{\kev}{\ensuremath{\,\mathrm{keV}}}
\newc{\mev}{\ensuremath{\,\mathrm{MeV}}}
\newc{\gev}{\ensuremath{\,\mathrm{GeV}}}
\newc{\tev}{\ensuremath{\,\mathrm{TeV}}}
\newc{\MeV}{\mev} 
\newc{\TeV}{\tev}
\newc{\invpb}{\ensuremath{/\text{pb}}}
\newc{\invfb}{\ensuremath{\,\text{fb}^{-1}}}
\newc\nb{\ensuremath{\,\mathrm{nb}}} \newc\pb{\ensuremath{\,\mathrm{pb}}} \newc\fb{\ensuremath{\,\mathrm{fb}}}
\newc\pc{\ensuremath{\,\mathrm{pc}}}
\newc\kpc{\ensuremath{\,\mathrm{kpc}}}
\newc\mpc{\ensuremath{\,\mathrm{Mpc}}}
\newc\ps{\ensuremath{\,\mathrm{ps}}} 
\newc\cmeter{\ensuremath{\,\mathrm{cm}}} 
\newc\meter{\ensuremath{\,\mathrm{m}}} 
\newc\kmeter{\ensuremath{\,\mathrm{km}}}
\newc\second{\ensuremath{\,\mathrm{s}}}
\newc\msecond{\ensuremath{\,\mathrm{ms}}}
\newc\nsecond{\ensuremath{\,\mathrm{ns}}}
\newc\psecond{\ensuremath{\,\mathrm{ps}}}
\newc{\chisqmin}{\ensuremath{\chi^2_{\mathrm{min}}}}
\newc{\Delchisq}{\ensuremath{\Delta\chi^2}}
\newc{\chisq}{\ensuremath{\chi^2}}
\newc{\like}{\ensuremath{\mathcal{L}}}
\newc\lsim{\ensuremath{\mathrel{\rlap{\lower4pt\hbox{\hskip1pt$\sim$}}\raise1pt\hbox{$<$}}}}
\newc\gsim{\ensuremath{\mathrel{\rlap{\lower4pt\hbox{\hskip1pt$\sim$}}\raise1pt\hbox{$>$}}}}
\newc{\VEV}[1]{\ensuremath{\langle #1 \rangle}}
\newc{\dl}{\ensuremath{\stackrel{\leftarrow}{D}}}
\newc{\dr}{\ensuremath{\stackrel{\rightarrow}{D}}}
\newc{\bcenter}{\begin{center}}    \newc{\ecenter}{\end{center}}
\newc{\bfl}{\begin{flushleft}}    \newc{\efl}{\end{flushleft}}
\newc{\bfr}{\begin{flushright}}    \newc{\efr}{\end{flushright}}

\newc{\bi}{\begin{itemize}}
\newc{\ei}{\end{itemize}}
\newc{\bed}{\begin{description}}
\newc{\eed}{\end{description}}
\newc{\ben}{\begin{enumerate}}
\newc{\een}{\end{enumerate}}

\newc{\be}{\begin{equation}}
\newc{\ee}{\end{equation}}
\newc{\bea}{\begin{eqnarray}}
\newc{\eea}{\end{eqnarray}}
\newc{\ra}{\rightarrow}
\newc{\alphas}{\ensuremath{\alpha_s}}
\newc{\alphatwo}{\ensuremath{\alpha_2}}
\newc{\alphaone}{\ensuremath{\alpha_1}}
\newc{\alphai}[1]{\ensuremath{\alpha_{#1}}}
\newc{\alphaem}{\ensuremath{\alpha_{\mathrm{em}}}}
\newc{\alphaeff}{\ensuremath{\alpha_{\mathrm{eff}}}}
\newc{\sineff}{\ensuremath{\sin \theta_{\mathrm{eff}}}}
\newc{\sinsqeff}{\ensuremath{\sin^2 \theta_{\mathrm{eff}}}}
\newc{\dalphahad}{\ensuremath{\Delta \alpha_{\mathrm{had}}}}
\newc{\yt}{\ensuremath{h_t}} \newc{\yb}{\ensuremath{h_b}} \newc{\ytau}{\ensuremath{h_{\tau}}}
\newc\mz{\ensuremath{M_Z}} 
\newc\mw{\ensuremath{m_W}}
\newc\mZ{\mz}        \newc\mW{\mw}
\newc\mhsm{\ensuremath{ m_{H_{\mathrm{SM}}}}}
\newc{\mtop}{\ensuremath{ m_t}}               \newc{\mtpole}{\ensuremath{ M_t}}
\newc{\mbottom}{\ensuremath{ m_b}} 
\newc{\mtau}{\ensuremath{ m_{\tau}}}
\newc{\mt}{\mtpole}
\newc{\mb}{\mbottom} 
\newc{\rgg}{\ensuremath{R_{h}(\gamma\gamma)}}
\newc{\rzz}{\ensuremath{R_{h}(ZZ)}}
\newc{\rtwogg}{\ensuremath{R_{h_2}(\gamma\gamma)}}
\newc{\rtwozz}{\ensuremath{R_{h_2}(ZZ)}}
\newc{\ronegg}{\ensuremath{R_{h_1}(\gamma\gamma)}}
\newc{\ronezz}{\ensuremath{R_{h_1}(ZZ)}}
\newc{\rsiggg}{\ensuremath{R_{h_\textrm{sig}}(\gamma\gamma)}}
\newc{\rsigzz}{\ensuremath{R_{h_\textrm{sig}}(ZZ)}}
\newc{\llbar}{\ensuremath{\ell\bar{\ell}}}
\newc{\tauptaum}{\ensuremath{ \tau^+\tau^-}}
\newc{\qqbar}{\ensuremath{ q\bar{q}}} \newc{\ppbar}{\ensuremath{ p\bar{p}}}
\newc{\bbbar}{\ensuremath{ b\bar{b}}} \newc{\ttbar}{\ensuremath{ t\bar{t}}}
\newc{\ffbar}{\ensuremath{ f\bar{f}}} \newc{\tautaubar}{\ensuremath{ \tau\bar{\tau}}}
\newc{\mchi}{\ensuremath{m_{\chi}}}
\newc{\squark}{\ensuremath{\tilde{q}}}
\newc{\slepton}{\ensuremath{\tilde{l}}}
\newc{\gluino}{\ensuremath{\tilde{g}}} 
\newc{\wino}{\ensuremath{\tilde{W}}}
\newc{\bino}{\ensuremath{\tilde{B}}}
\newc{\mgluino}{\ensuremath{{m_{\gluino}}}}
\newc{\tone}{\ensuremath{{\tilde{t}_1}}}
\newc{\sthw}{\ensuremath{ \sin\theta_W}}              \newc{\cthw}{\ensuremath{\cos\theta_W}}
\newc{\tanthw}{\ensuremath{ \tan\theta_W}}              \newc{\cotthw}{\ensuremath{\cot\theta_W}}
\newc{\ssqthw}{\ensuremath{\sin^2 \theta_W}}
\newc{\msbar}{\ensuremath{\overline{MS}}} \newc{\drbar}{\ensuremath{\overline{DR}}}
\newc{\mtmtsmmsbar}{\ensuremath{ m_t(m_t)^{\msbar}_{{\mathrm{SM}}}}}
\newc{\mtmtsmdrbar}{\ensuremath{ m_t(m_t)^{\drbar}_{{\mathrm{SM}}}}}
\newc{\mtmtmssmdrbar}{\ensuremath{ m_t(m_t)^{\drbar}_{{\mathrm{SUSY}}}}}
\newc{\mbmbmsbar}{\ensuremath{ m_b(m_b)^{\msbar} }}
\newc{\mbmbsmmsbar}{\ensuremath{ m_b(m_b)^{\msbar}_{{\mathrm{SM}}}}}
\newc{\mbmzsmmsbar}{\ensuremath{ m_b(\mz)^{\msbar}_{{\mathrm{SM}}}}}
\newc{\mbmzsmdrbar}{\ensuremath{ m_b(\mz)^{\drbar}_{{\mathrm{SM}}}}}
\newc{\mbmzmssmdrbar}{\ensuremath{ m_b(\mz)^{\drbar}_{{\mathrm{SUSY}}}}}
\newc{\mtaumzsmmsbar}{\ensuremath{ m_{\tau}(\mz)^{\msbar}_{{\mathrm{SM}}}}}
\newc{\mtaumzsmdrbar}{\ensuremath{ m_{\tau}(\mz)^{\drbar}_{{\mathrm{SM}}}}}
\newc{\mtaumzmssmdrbar}{\ensuremath{ m_{\tau}(\mz)^{\drbar}_{{\mathrm{SUSY}}}}}
\newc{\alphasmzms}{\ensuremath{\alpha_s(M_Z)^{\overline{MS}}}}
\newc{\alphaimzms}[1]{\ensuremath{\alpha_{#1}(M_Z)^{\overline{MS}}}}
\newc{\alphaemmz}{\ensuremath{\alpha_{\mathrm{em}}(M_Z)^{\overline{MS}}}}
\newc{\mzero}{\ensuremath{{m_0}}}
\newc{\mhalf}{\ensuremath{ m_{1/2}}}
\newc{\tanb}{\ensuremath{\tan\beta}}
\newc{\azero}{\ensuremath{ A_0}}
\newc{\bzero}{\ensuremath{ B_0}}
\newc{\signmu}{\ensuremath{\rm{sgn}\,\mu}}
\newc{\mueff}{\ensuremath{\mu_{\rm{eff}}}}
\newc{\lam}{\ensuremath{{\lambda}}}
\newc{\kap}{\ensuremath{{\kappa}}}
\newc{\alam}{\ensuremath{{A_{\lambda}}}}
\newc{\akap}{\ensuremath{{A_{\kappa}}}}
\newc{\hs}{\ensuremath{ H_s}}      
\newc{\mhs}{\ensuremath{ m_{H_s}}} 
\newc{\mgut}{\ensuremath{ M_{\rm GUT}}}
\newc{\mplanck}{\ensuremath{ M_{\rm P}}}      \newc{\mpl}{\ensuremath{ M_{\rm Pl}}}
\newc{\msusy}{\ensuremath{ M_{\rm SUSY}}}      \newc{\ms}{\ensuremath{ M_{\rm S}}}
 \newc{\mhl}{\ensuremath{m_\hl}} 
 \newc{\mhone}{\ensuremath{m_{h_1}}} 
 \newc{\mhtwo}{\ensuremath{m_{h_2}}} 
 \newc{\mglu}{\ensuremath{m_{\tilde g}}} 
 \newc{\mul}{\ensuremath{m_{\tilde{u}_L}}} 
 \newc{\mtone}{\ensuremath{m_{\tilde{t}_1}}} 
 \newc{\ma}{\ensuremath{m_A}} 
 \newc{\maone}{\ensuremath{m_{a_1}}} 
 \newc{\matwo}{\ensuremath{m_{a_2}}}
 \newc{\hone}{\ensuremath{h_1}}
 \newc{\htwo}{\ensuremath{h_2}}
 \newc{\aone}{\ensuremath{a_1}}
 \newc{\atwo}{\ensuremath{a_2}}
 \newc{\mhu}{\ensuremath{ m_{H_u}}}       
 \newc{\mhd}{\ensuremath{ m_{H_d}}}
 \newc{\mhusq}{\ensuremath{ m_{H_u}^2}}       
 \newc{\mhdsq}{\ensuremath{ m_{H_d}^2}}
 \newc{\mhuew}{\ensuremath{ m^{\ast}_{H_u}}}       
 \newc{\mhdew}{\ensuremath{ m^{\ast}_{H_d}}}
 \newc{\mhuewsq}{\ensuremath{ m^{\ast\, 2}_{H_u}}}       
 \newc{\mhdewsq}{\ensuremath{ m^{\ast\, 2}_{H_d}}}
 \newc{\hu}{\ensuremath{ H_u}}       
 \newc{\hd}{\ensuremath{ H_d}}
 \newc{\barmhu}{\ensuremath{ \bar{m}_{H_u}}}
 \newc{\barmhd}{\ensuremath{ \bar{m}_{H_d}}}

 \newc{\mqthree}{\ensuremath{m_{\widetilde{Q}_3}^2}}
 \newc{\muthree}{\ensuremath{m_{\tilde{u}_3}^2}}
 \newc{\mdthree}{\ensuremath{m_{\tilde{d}_3}^2}}
 \newc{\mlthree}{\ensuremath{m_{\widetilde{L}_3}^2}}
 \newc{\methree}{\ensuremath{m_{\tilde{e}_3}^2}}
 \newc{\mqtwo}{\ensuremath{m_{\widetilde{Q}_2}^2}}
 \newc{\mutwo}{\ensuremath{m_{\tilde{u}_2}^2}}
 \newc{\mdtwo}{\ensuremath{m_{\tilde{d}_2}^2}}
 \newc{\mltwo}{\ensuremath{m_{\widetilde{L}_2}^2}}
 \newc{\metwo}{\ensuremath{m_{\tilde{e}_2}^2}}
 \newc{\mqone}{\ensuremath{m_{\widetilde{Q}_1}^2}}
 \newc{\muone}{\ensuremath{m_{\tilde{u}_1}^2}}
 \newc{\mdone}{\ensuremath{m_{\tilde{d}_1}^2}}
 \newc{\mlone}{\ensuremath{m_{\widetilde{L}_1}^2}}
 \newc{\meone}{\ensuremath{m_{\tilde{e}_1}^2}}
 \newc{\mone}{\ensuremath{M_1}}
 \newc{\monesq}{\ensuremath{M_1^2}}
 \newc{\mtwo}{\ensuremath{M_2}}
 \newc{\mtwosq}{\ensuremath{M_2^2}}
 \newc{\mthree}{\ensuremath{M_3}}
 \newc{\mthreesq}{\ensuremath{M_3^2}}
 \newc{\atau}{\ensuremath{{A_{\tau}}}}
 \newc{\at}{\ensuremath{{A_{t}}}}
 \newc{\ab}{\ensuremath{{A_{b}}}}
 \newc{\atausq}{\ensuremath{{A_{\tau}^2}}}
 \newc{\atsq}{\ensuremath{{A_{t}^2}}}
 \newc{\absq}{\ensuremath{{A_{b}^2}}}
 \newc{\dmzero}{\ensuremath{\Delta{_{m_0}}}}
 \newc{\dmhalf}{\ensuremath{\Delta{_{m_{1/2}}}}}
 \newc{\dmu}{\ensuremath{\Delta{_{\mu}}}}

 \newc{\pten}{\ensuremath{\psi_{10}}}
 \newc{\ffive}{\ensuremath{\phi_{5}}}
 \newc{\hfive}{\ensuremath{h_{5}}}
 \newc{\hbfive}{\ensuremath{h_{\bar{5}}}}
 \newc{\thet}{\ensuremath{\theta_{50}}}
 \newc{\thetb}{\ensuremath{\theta_{\,\overline{50}}}}
 \newc{\ptenhat}{\ensuremath{\hat{\psi}_{10}}}
 \newc{\ffivehat}{\ensuremath{\hat{\phi}_{5}}}
 \newc{\hfivehat}{\ensuremath{\hat{h}_{5}}}
 \newc{\hbfivehat}{\ensuremath{\hat{h}_{\bar{5}}}}
 \newc{\thethat}{\ensuremath{\hat{\theta}_{50}}}
 \newc{\thetbhat}{\ensuremath{\hat{\theta}_{\,\overline{50}}}}
 \newc{\si}{\ensuremath{\Sigma}}
 \newc{\mfive}{\ensuremath{m_5^2}}
 \newc{\mten}{\ensuremath{m_{10}^2}}
 \newc{\dfive}{\ensuremath{\Delta^2_5}}
 \newc{\dbfive}{\ensuremath{\Delta^2_{\bar{5}}}}
 \newc{\dfifty}{\ensuremath{\Delta^2_{50}}}
 \newc{\dfiftyb}{\ensuremath{\Delta^2_{\,\overline{50}}}}
 \newc{\msi}{\ensuremath{m_{\Sigma}^2}}
 \newc{\lamh}{\ensuremath{\lambda_{H}}}
 \newc{\lamhb}{\ensuremath{\lambda_{\bar{H}}}}
 \newc{\ah}{\ensuremath{A_{H}}}
 \newc{\ahb}{\ensuremath{A_{\bar{H}}}}
 \newc{\lams}{\ensuremath{\lambda_{S}}}
 \newc{\as}{\ensuremath{A_{S}}}
 \newc{\lamsig}{\ensuremath{\lambda_{\si}}}
 \newc{\asig}{\ensuremath{A_{\si}}}

 \newc{\msten}{\ensuremath{m_{16}^2}}
 \newc{\mhun}{\ensuremath{m_{126}^2}}
 \newc{\mhunb}{\ensuremath{m_{\bar{126}}^2}}
 \newc{\mthun}{\ensuremath{m_{210}^2}}
 \newc{\ahun}{\ensuremath{A_{\bar{126}}}}
 \newc{\yhun}{\ensuremath{Y_{\bar{126}}}}
 \newc{\aten}{\ensuremath{A_{10}}}
 \newc{\yten}{\ensuremath{Y_{10}}}
 \newc{\alone}{\ensuremath{A_{\lambda_1}}}
 \newc{\altwo}{\ensuremath{A_{\lambda_2}}}
 \newc{\althree}{\ensuremath{A_{\lambda_3}}}
 \newc{\althreeb}{\ensuremath{A_{\bar{\lambda_3}}}}
 \newc{\lone}{\ensuremath{\lambda_1}}
 \newc{\ltwo}{\ensuremath{\lambda_2}}
 \newc{\lthree}{\ensuremath{\lambda_3}}
 \newc{\lthreeb}{\ensuremath{\bar{\lambda_3}}}

\newc{\sigsip}{\ensuremath{\sigma^{\rm SI}_{p}}}	\newc{\sigsin}{\ensuremath{\sigma^{\rm SI}_{n}}}
\newc{\sigsdp}{\ensuremath{\sigma^{\rm SD}_{p}}}	\newc{\sigsdn}{\ensuremath{\sigma^{\rm SD}_{n}}}
\newc{\sigsi}{\ensuremath{\sigma^{\rm SI}}}	\newc{\sigsd}{\ensuremath{\sigma^{\rm SD}}}
\newc{\sigv}{\ensuremath{\sigma v}}
\newc{\abund}{\ensuremath{ \Omega h^2}}
\newc{\omegadm}{\ensuremath{ \Omega_{{\rm DM}}}}     \newc{\abunddm}{\ensuremath{ \Omega_{{\rm DM}} h^2}} 
\newc{\omegam}{\ensuremath{ \Omega_{{\rm m}}}}       \newc{\abundm}{\ensuremath{ \Omega_{{\rm m}} h^2}}
\newc{\omegab}{\ensuremath{ \Omega_{{\rm b}}}}	\newc{\abundb}{\ensuremath{ \Omega_{{\rm b}} h^2}}
\newc{\omegatot}{\ensuremath{ \Omega_{{\rm TOT}}}}
\newc{\omegacdm}{\ensuremath{ \Omega_{{\rm CDM}}}}   \newc{\abundcdm}{\ensuremath{ \Omega_{{\rm CDM}} h^2}}
\newc{\omegalambda}{\ensuremath{ \Omega_{\Lambda}}} \newc{\abundlambda}{\ensuremath{ \Omega_{\Lambda} h^2}}
\newc{\omegarad}{\ensuremath{ \Omega_{{\rm rad}}}}  \newc{\abundrad}{\ensuremath{ \Omega_{{\rm rad}} h^2}}
\newc{\rhocrit}{\ensuremath{ \rho_{\rm crit}}}
\newc{\rhochi}{\ensuremath{ \rho_{\chi}}}
\newc{\abunchi}{\ensuremath{\Omega_\chi h^2}}
\newc{\abundlsp}{\ensuremath{\Omega_{\rm LSP}h^2}}
\newc{\abundchi}{\ensuremath{\Omega_\chi h^2}}% For multiple citations
\newc{\tf}{\ensuremath{T_f}} \newc{\xf}{\ensuremath{x_f}}
\newc{\tr}{\ensuremath{T_R}}

\newc{\amu}{\ensuremath{ a_{\mu}}}        
\newc{\amususy}{\ensuremath{ a_{\mu}^{\mathrm{SUSY}}}}
\newc{\amuexpt}{\ensuremath{ a_{\mu}^{\mathrm{expt}}}}        
\newc{\amusm}{\ensuremath{ a_{\mu}^{\mathrm{SM}}}}
\newc\deltaamu{\ensuremath{\delta a_{\mu}}} 
\newc{\deltaamususy}{\ensuremath{\delta a_{\mu}^{\mathrm{SUSY}}}}
\newc\gmtwo{\ensuremath{ (g-2)_{\mu}}} 
\newc{\deltagmtwomususy}{\ensuremath{\delta\left(g-2\right)_{\mu}^{\mathrm{SUSY}}}}
\newc{\deltagmtwomu}{\ensuremath{\delta\left(g-2\right)_{\mu}}}

\newc{\amubnl}{\ensuremath{a_{\mu}^{\mathrm{BNL}}}}
\newc{\amufermi}{\ensuremath{a_{\mu}^{\mathrm{FNAL}}}}
\newc{\amufermicombo}{\ensuremath{a_{\mu}^{\mathrm{FNAL+BNL}}}}
\newc{\deltaamubnl}{\ensuremath{\delta a_{\mu}^{\mathrm{BNL}}}}
\newc{\deltaamufermi}{\ensuremath{\delta a_{\mu}^{\mathrm{FNAL}}}}
\newc{\deltaamufermicombo}{\ensuremath{\delta a_{\mu}^{\mathrm{FNAL+BNL}}}}

\newc{\gmtwobnl}{\ensuremath{\left(g-2\right)_{\mu}^{\mathrm{BNL}}}}
\newc{\gmtwofermi}{\ensuremath{\left(g-2\right)_{\mu}^{\mathrm{FNAL}}}}

\newc{\deltagmtwobnl}{\ensuremath{\delta\left(g-2\right)_{\mu}^{\mathrm{BNL}}}}
\newc{\deltagmtwofermi}{\ensuremath{\delta\left(g-2\right)_{\mu}^{\mathrm{FNAL}}}}

\newc\BR{\ensuremath{\rm BR}}
\newc\bsgamma{\ensuremath{ b\rightarrow s \gamma }}
\newc\bxsgamma{\ensuremath{\overline{B}\rightarrow X_{s}\gamma}}
\newc\brbsgamma{\ensuremath{\BR\left(\bsgamma\right)}}
\newc\brbxsgamma{\ensuremath{\BR\left(\bxsgamma\right)}}
\newc\bsmumu{\ensuremath{B_s\to\mu^+\mu^-}}
\newc\brbsmumu{\ensuremath{\BR\left(B_s\to\mu^+\mu^-\right)}}
\newc\bdmmumu{\ensuremath{\overline{B}_d\to\mu^+\mu^-}}
\newc\bbbarmix{\ensuremath{\overline{B}_s\mbox{-}B_s}}      % B_s mixing
\newc\delmbs{\ensuremath{\Delta M_{B_s}}}
\newc{\butaunu}{\ensuremath{B_u \rightarrow \tau \nu}}
\newc{\brbutaunu}{\ensuremath{\BR\left(B_u \rightarrow \tau \nu\right)}}
\newcommand*{\charone}{\ensuremath{\chi^{\pm}_1}}

 \newc{\msmul}{\ensuremath{m_{\tilde{\mu}_L}}}
\newc{\msmur}{\ensuremath{m_{\tilde{\mu}_R}}}
\newc{\msneumu}{\ensuremath{m_{\tilde{\nu}_{\mu}}}}

\let\oldcite\cite
\renewcommand*{\cite}{~\oldcite}

\newcommand*{\hl}{\ensuremath{h}}

\usepackage[utf8]{inputenc}
\usepackage{amsmath,amssymb,amsthm,amsfonts}
\usepackage{graphicx}
\usepackage{xspace}
\usepackage{subcaption}
\usepackage{dcolumn}	% Align table columns on decimal point
\usepackage{hyperref}      	% hypertext links %%ARXIV
\usepackage{bm}		% bold math
\usepackage{epsfig}
\usepackage{epstopdf}
\usepackage{setspace}
\usepackage[usenames, dvipsnames]{color}
\usepackage{slashed}
\usepackage{comment}
\usepackage{enumitem}
\usepackage{array,multirow}
\usepackage{feynmp}
\usepackage{datetime}
\usepackage{enumitem}
\usepackage{titlesec, titletoc}
\usepackage{multirow}
\titleformat*{\section}{\boldmath\bfseries}
\titleformat*{\subsection}{\boldmath\bfseries}
\setlist[description]{leftmargin=0.4cm}
\makeatletter
\def\endfmffile{%
	\fmfcmd{\p@rcent\space the end.^^J%
		end.^^J%
		endinput;}%
	\if@fmfio
	\immediate\closeout\@outfmf
	\fi
	\ifnum\pdfshellescape>\z@
	\immediate\write18{mpost \thefmffile}%
	\fi}
\makeatother
\RequirePackage[normalem]{ulem}

\count\footins = 1000

\begin{document}
\thispagestyle{empty}
\def\thefootnote{\fnsymbol{footnote}}

\begin{center}

{\large \bf {GUT-constrained supersymmetry and dark matter in light of the new {\boldmath $(g-2)_\mu$} determination}}

\vspace{1cm}

{ Manimala Chakraborti$^{1}$%
  \footnote{email: mani.chakraborti@gmail.com}%
  , Leszek Roszkowski$^{1,2}$%
  \footnote{email: leszek.roszkowski@ncbj.gov.pl}%
  , Sebastian Trojanowski$^{1,2}$%
  \footnote{email: strojanowski@camk.edu.pl}%
}

\vspace*{.7cm}

{\it $^1$ Astrocent, Nicolaus Copernicus Astronomical Center Polish                                                                                                                                          Academy of Sciences, \\                                                                                                                                                                                     ul.~Rektorska 4, 00-614, Warsaw, Poland\\                                                                                                                                                                
\vspace*{0.2cm}                                                                                                         
$^2$ National Centre for Nuclear Research, \\                                                          ul.~Pasteura 7, 02-093 Warsaw, Poland}

\end{center}

\vspace*{0.1cm}

\def\thefootnote{\arabic{footnote}}
\setcounter{page}{0}
\setcounter{footnote}{0}

\begin{abstract}
%\PRE{\vspace*{0.2in}}
\vspace*{0.1in}
The recent confirmation by the Fermilab-based Muon g-2 experiment of the  \gmtwo\ anomaly  has important implications for allowed particle spectra in softly broken supersymmetry (SUSY) models with neutralino dark matter (DM). Generally, the DM has to be quite light,  with the mass up to a few hundred GeV, and bino-dominated if it is to provide most of DM in the Universe. Otherwise, a higgsino or wino dominated DM is also allowed but only as a strongly subdominant component of at most a few percent of the total density. These general patterns can easily be found in the  phenomenological models of SUSY but in GUT-constrained scenarios this proves much more challenging. In this paper we revisit the issue in the framework of some unified SUSY models with different GUT boundary conditions on the soft masses. We study the so-called non-universal gaugino model (NUGM) in which the mass of the gluino is disunified from those of the bino and the wino and an $SO(10)$ and an $SU(5)$ GUT-inspired models as examples. We find that in these unified frameworks the above two general patterns of DM can also be found, and thus the muon anomaly can also be accommodated, unlike in the simplest frameworks of the CMSSM or the NUHM. We show the resulting values of direct detection cross-section for points that do and do  not satisfy the muon anomaly. On the other hand, it will be challenging to access those solutions at the LHC because the resulting spectra are generally very compressed.
\end{abstract}

%\pacs{}

%\pagenumbering{gobble}
\maketitle
%\thispagestyle{empty}

%\renewcommand{\baselinestretch}{0.95}\normalsize
%\tableofcontents
%\renewcommand{\baselinestretch}{1.0}\normalsize

%\pagenumbering{arabic}

%\clearpage

%%%%%%%%%%%%%%%%%%%%%%%%%%%%%%%%%%%%%%%%%%%%%%%%%%%%%%%%%%%%%%%%
\section{\label{sec:intro}Introduction}
%%%%%%%%%%%%%%%%%%%%%%%%%%%%%%%%%%%%%%%%%%%%%%%%%%%%%%%%%%%%%&&&&

The  persistent discrepancy of the determination of the anomalous magnetic moment of the muon $\amu \equiv \gmtwo/2$ between the experimental measurement of the Brookhaven experiment E821\cite{Bennett:2006fi} and the Standard Model (SM) value\cite{Aoyama:2020ynm} has arguably been  the {\em only}  significant data-based argument pointing towards a sub-TeV scale of beyond the Standard Model (BSM) physics. 
The discrepancy 
$\deltaamubnl = \amubnl - \amusm = (28.1 \pm 7.6) \times10^{-10} $ 
of about $3.7\sigma$  has remained at roughly that level for several years amid various improvements and updates in the various contributions to the SM value. 
For a recent review, see Ref.\cite{Aoyama:2020ynm} which summarises a large volume of work on the subject.
An eagerly expected new measurement by the Fermilab-based Muon g-2 experiment has just been announced that, when combined with the BNL result gives\cite{Abi:2021gix}
\be
\deltaamu = \amufermicombo - \amusm = (25.1 \pm 5.9) \times10^{-10},
\label{eq:fnalgmtwo}
\ee
which gives a $4.2\sigma$ discrepancy with the SM value\footnote{It is, however, worth stressing that recently a lattice computation of the leading order SM hadronic vaccuum polarization (LO-HVP) contribution has appeared\cite{Borsanyi:2020mff} which largely reduces the tension between the theoretical and experimental results.}. 
The result can easily be explained in a number of BSM scenarios but only for very specific patterns and choices of parameters. In this sense the \gmtwo\ result has far reaching consequences for our projections for BSM physics as a strong BSM pattern discriminator.

It is well known that current LHC limits still easily allow SUSY corrections to \deltaamu\  to  be quite large but only in phenomenological frameworks of SUSY with all parameters defined at the electroweak (EW) scale, like the Minimal Supersymmetric Standard Model (MSSM)\cite{Fowlie:2013oua,Chakraborti:2014gea,Choudhury:2016lku,Chakraborti:2017vxz,Bagnaschi:2017tru,Hagiwara:2017lse,Cox:2018qyi,Carena:2018nlf,Abdughani:2019wai,Endo:2020mqz,Chakraborti:2020vjp,Chakraborti:2021kkr}. The situation is generally very different in unified, i.e., based on grand unification theories (GUTs), or -- in other words -- GUT-constrained, SUSY models, where electroweakly interacting states (EW-inos) are usually tied up to the coloured sector and often cannot be arbitrarily light. 

The level of  contribution to \deltaamu\ clearly depends on the spectrum of the model at hand. In a simple framework of the Constrained MSSM (CMSSM) the contribution to \gmtwo\ is very small and it can reproduce neither the previous nor the new value of \gmtwo; see, e.g.,\cite{Cao:2011sn,Bechtle:2012zk,Fowlie:2012im,Buchmueller:2012hv,Strege:2012bt,Bechtle:2015nua,Han:2016gvr,Athron:2017qdc}.  This is so due to the combination of direct lower limits on coloured sparticles at the LHC
and the mass of the Higgs boson which both push the favoured parameter space for the unified masses of scalars and gauginos, \mzero\ and \mhalf, respectively, to the multi-\tev\ regime, as stated above. This prevents light sleptons, binos, and winos from being light enough.

It has been known that this pattern can be easily relaxed in other GUT-based models with less stringent unification assumptions. 
For instance, as was initially shown in\cite{Akula:2013ioa}, this can be achieved when the condition of gaugino unification is relaxed. In this case, when the high-scale value of the gluino soft mass, $M_3$, is much larger than the values of the bino and wino masses, \mone\ and \mtwo, respectively, the renormalization group equations (RGEs) can drive coloured sparticle masses to values of a few-TeV, or even larger, consistent with the LHC limits. Instead, EW-inos are allowed to be light enough to contribute to $\deltaamu$\cite{Kowalska:2015kaa,Kowalska:2015zja}; see also\cite{Mohanty:2013soa,Chakrabortty:2013voa,Gogoladze:2014cha,Ajaib:2015ika,Cox:2018vsv,Tran:2018kxv}.

In this paper we revisit the situation in some GUT-constrained SUSY models in light of the new data from the Fermilab Muon g-2 experiment\footnote{The impact of the new muon g-2  result on the parameter space of SUSY models has been discussed in several recent works, see e.g.\cite{Abdughani:2021pdc,Cao:2021tuh,Chakraborti:2021dli,Ibe:2021cvf,Han:2021ify,wang2021gutscale,Heinemeyer:2021zpc,iwamoto2021winohiggsino,Baum:2021qzx,Yin:2021mls,VanBeekveld:2021tgn,Gu:2021mjd,Cox:2021gqq,Endo:2021zal,Zhang:2021gun,Yang:2021duj}. A very recent exhaustive review on the topic can be found in\cite{Athron:2021iuf}.} .
We focus in particular on the models that allow light enough EW-inos\cite{Kowalska:2015zja}. As an additional ingredient, we require that the lightest superpartner (LSP) is neutral. We will examine when its relic density is consistent with cosmic determinations of dark matter (DM)\cite{Aghanim:2018eyx}. In unified models this condition is typically very constraining, especially at the low LSP mass in a few hundred GeV regime implied by the \deltaamu\ result, and allowing for bino-like LSP only.
For this reason we will also examine cases when the LSP can be only a subdominant component of DM. This will allow for light higgsino-like and wino-like cases as well.

We will examine in particular some models that not only have the potential to resolve the \gmtwo\ anomaly but in addition show less fine tuning than the simplest unified models, like the CMSSM.
Specific unification patterns and mass relations can lead to a significant lowering of the fine tuning due to gauginos, scalars, and the $\mu$ parameter, relative to the simplest unification conditions\cite{Kowalska:2014hza}. In particular, for a nearly pure higgsino with mass of about $1\tev$  the fine tuning can be reduced to the level of a few percent, while the scalars and gauginos have masses in the multi-TeV regime. 

The paper is organised as follows. In Sec.~\ref{sec:generalremarks} we give an estimate of 
SUSY contribution to \gmtwo\ for MSSM scenarios featuring light higgsinos in the spectrum. In Sec.~\ref{sec:models} we describe the specific GUT-constrained SUSY scenarios considered in this work. Our numerical setup and parameter scanning procedure is discussed in Sec.~\ref{sec:simulation}. In Sec.~\ref{sec:results} we give a description of our results. Finally, we conclude in Sec.~\ref{sec:conc}.

%%%%%%%%%%%%%%%%%%%%%%%%%%%%%%%%%%%%%%%%%%%%%%%%%%%%%%%%%%%%%&&&&
\section{\label{sec:generalremarks}Supersymmetric contributions to $(g-2)_\mu$}
%%%%%%%%%%%%%%%%%%%%%%%%%%%%%%%%%%%%%%%%%%%%%%%%%%%%%%%%%%%%%&&&&
Dominant SUSY contribution to $\gmtwo$ at one loop comes from diagrams involving
chargino-sneutrino and neutralino-smuon in the loop. In the mass insertion
approximation the five main contributions can be given as\cite{Moroi:1995yh,Martin:2001st}

\be
\Delta_{\charone \tilde{\nu}_{\mu}}=\frac{g^2}{(4 \pi)^2}\frac{m_{\mu}^2\tanb}{\mu\mtwo}\,\mathcal{F}_{[\charone \tilde{\nu}_{\mu}]}\left(\frac{\mu^2}{\msneumu^2},\frac{\mtwo^2}{\msneumu^2}\right)\,,
\label{charsneu}
\ee
\be
\Delta^{(1)}_{\chi\, \tilde{\mu}}=-\frac{1}{2}\frac{g^2}{(4 \pi)^2}\frac{m_{\mu}^2\tanb}{\mu\mtwo}\,\mathcal{F}_{[\chi\, \tilde{\mu}]}\left(\frac{\mu^2}{\msmul^2},\frac{\mtwo^2}{\msmul^2}\right)\,,
\label{neusmu1}
\ee
\be
\Delta^{(2)}_{\chi\, \tilde{\mu}}=\frac{1}{2}\frac{g'^2}{(4 \pi)^2}\frac{m_{\mu}^2\tanb}{\mu\mone}\,\mathcal{F}_{[\chi\, \tilde{\mu}]}\left(\frac{\mu^2}{\msmul^2},\frac{\mone^2}{\msmul^2}\right)\,,
\label{neusmu2}
\ee
\be
\Delta^{(3)}_{\chi\, \tilde{\mu}}=-\frac{g'^2}{(4 \pi)^2}\frac{m_{\mu}^2\tanb}{\mu\mone}\,\mathcal{F}_{[\chi\, \tilde{\mu}]}\left(\frac{\mu^2}{\msmur^2},\frac{\mone^2}{\msmur^2}\right)\,,
\label{neusmu3}
\ee
\be
\Delta^{(4)}_{\chi\, \tilde{\mu}}=\frac{g'^2}{(4 \pi)^2}\frac{m_{\mu}^2\mone\mu}{\msmul^2 \msmur^2}\tanb\,\mathcal{F}_{[\chi\, \tilde{\mu}]}\left(\frac{\msmur^2}{\mone^2},\frac{\msmul^2}{\mone^2}\right)\,,
\label{neusmu4}
\ee
where $g$ and $g'$ are the $SU(2)$ and $U(1)_Y$ SM gauge couplings respectively, and the loop functions $\mathcal{F}_{[\charone \tilde{\nu}_{\mu}]}$ and $\mathcal{F}_{[\chi\, \tilde{\mu}]}$ are defined as
\bea
\mathcal{F}_{[\charone \tilde{\nu}_{\mu}]}(x,y)&=& xy\left\{\frac{5-3(x+y)+xy}{(x-1)^2(y-1)^2}-\frac{2}{x-y}\left[\frac{\ln x}{(x-1)^3}-\frac{\ln y}{(y-1)^3}\right]\right\}\,,\\
\mathcal{F}_{[\chi\,\tilde{\mu}]}(x,y)&=& xy\left\{\frac{-3+x+y+xy}{(x-1)^2(y-1)^2}+\frac{2}{x-y}\left[\frac{x\ln x}{(x-1)^3}-\frac{y\ln y}{(y-1)^3}\right]\right\}\,.
\eea
In the limit of degenerate SUSY masses, the dominant contribution to $\gmtwo$ comes from the chargino-sneutrino loop of Eq.~\ref{charsneu}. This is true for scenarios involving  a light higgsino in the spectrum. As the value of $\mu$
increases, the contribution from neutralino-smuon loop, as given in Eq.~\ref{neusmu4} may become the dominant one provided both $\msmul$ and $\msmur$ are sufficiently small.

\begin{figure}%[htb!]                                                       
%\vspace{2em}
\centering
    \includegraphics[width=0.5\textwidth]{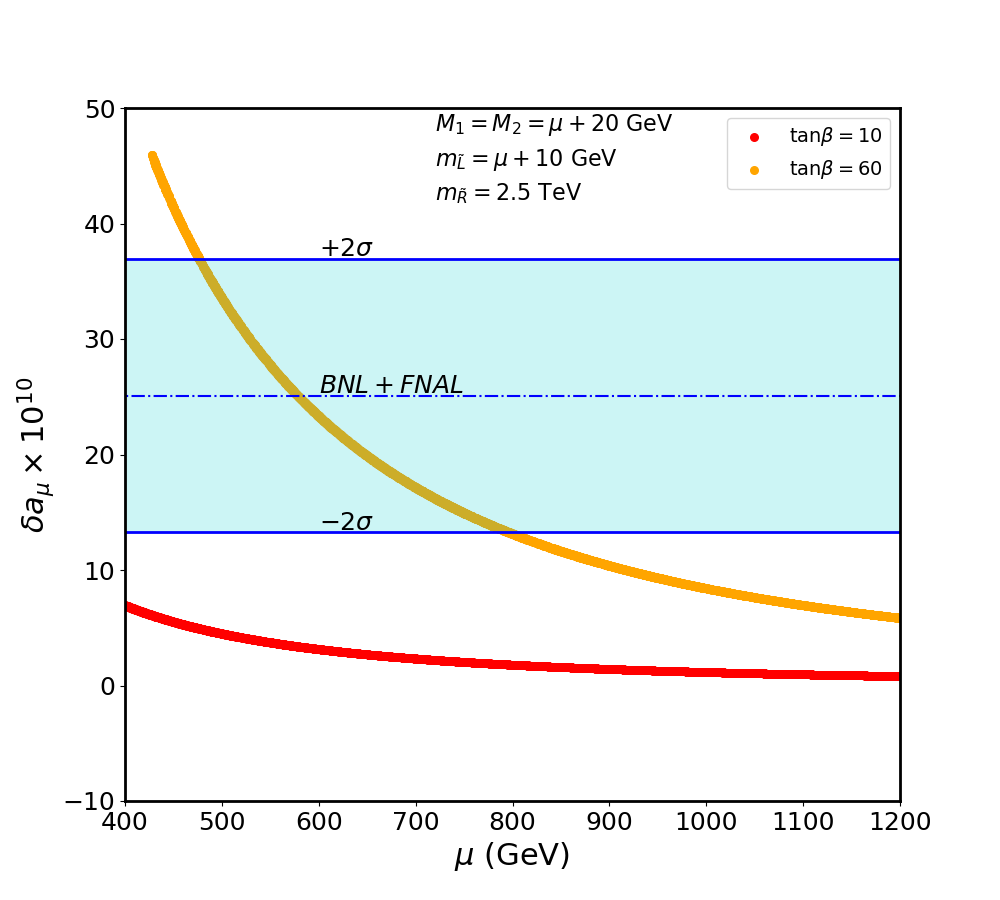}
\caption{\small The plots show the SUSY contributions to $\gmtwo$ as a function of $\mu$ for the mixed LSP scenario corresponding to  the parameter choice $M_1 = M_2 = \mu + 20 \gev$, $\msmul = \mu +10 \gev$, $\msmur =2.5 \tev$. The results are shown in red and orange for $\tanb = 10$ and $60$, respectively. The  central value and its $\pm 2\sigma$ limits for the deviation based on the combined experimental measurement (Eq.~\ref{eq:fnalgmtwo}) are shown as a cyan band.}
\label{fig:mixed}
\end{figure}

To give an estimate of the SUSY contributions to $\gmtwo$,
we plot in Fig.~\ref{fig:mixed} $\gmtwo$ as a function of $\mu$.
Fig.~\ref{fig:mixed} corresponds to the parameter choice $M_1 = M_2 = \mu + 20 \gev$.
This choice ensures that the LSP is predominantly higgsino-like with significant
wino and bino admixture in it. The smuon masses are chosen so as to maximize the
contribution from Eq.~\ref{charsneu}, namely,  $\msmul = \mu + 10 \gev$ and
$\tilde \mu_R$ is taken to be decoupled from the spectrum.  The relatively
small values of $\mu$ and the decoupled $\tilde \mu_R$ in this scenario
implies that the contribution from the neutralino-smuon loops in this
case is negligibly small. Thus, a significantly large contribution
to $\gmtwo$ requires the sneutrino mass to be sufficiently small,
which is ensured by this choice. It is observed that
the $\gmtwo$ anomaly can be explained for values of $m_\chi \sim \mu$
up to $\sim 900 \gev$  for $\tanb = 60$, considering the $2\sigma$
uncertainty on the central value.

%%%%%%%%%%%%%%%%%%%%%%%%%%% F I G U R E %%%%%%%%%%%%%%%%%%%%%%%%%%%%%%                                                                                  

%%%%%%%%%%%%%%%%%%%%%%%%%%%%%%%%%%%%%%%%%%%%%%%%%%%%%%%%%%%%%%%%%%%%%%
\section{\label{sec:GUT}GUT-constrained supersymmetry and $(g-2)_\mu$}
%%%%%%%%%%%%%%%%%%%%%%%%%%%%%%%%%%%%%%%%%%%%%%%%%%%%%%%%%%%%%%%%%%%%%%
\subsection{\label{sec:models}Selected GUT-constrained scenarios}

We perform our analysis in the framework of six specific GUT-constrained models that are described below. We  start with the most unified scenarios and then move to the less constrained ones in order to illustrate the impact of loosening the GUT constraints on the allowed range of $\deltaamu$. The relevant parameters and their ranges are given in Tables~\ref{tab:parameters} and \ref{tab:ranges}.

A prototype GUT-constrained scenario is the {\bf constrained MSSM (CMSSM)}\cite{Kane:1993td} in which all the soft scalar and gaugino masses are unified to their respective common parameters, $m_0$ and $m_{1/2}$. Similarly, all the trilinear couplings are taken as equal to a single $A_0$ parameter at the GUT scale and we will assume that the same is true for all the other models considered below. We also vary the ratio of the Higgs vacuum expectation values (vevs), $\tan\beta$, while we keep the sign of the $\mu$ parameter positive, $\textrm{sgn}(\mu)>0$. The last choice is favored by the sign of $\deltaamu$\footnote{
We keep the same choice of $\textrm{sgn}(\mu)$ also for less minimal scenarios discussed below. We note that, in general, the sign of the supersymmetric contribution to $\deltaamu$ is driven by $\textrm{sgn}(M_{1,2}\mu)$. When presenting our results below, we focus on scenarios predicting a positive BSM contribution to $\deltaamu$.}. As is well-known, already in this minimal GUT-constrained model, one can accommodate for the SM-like Higgs boson with the mass $m_h\simeq 125\gev$\cite{Fowlie:2012im} and simultaneously obtain the DM relic density either for the bino- or higgsino-like lightest neutralino\cite{Kowalska:2013hha}; cf. also Refs\cite{Bechtle:2012zk,Roszkowski:2014wqa,Athron:2017qdc,Athron:2017fxj} for other analyses.

An alternative, slightly less unified scenario, is to assume that the soft mass parameters in the Higgs sector, $m_{H_d}$ and $m_{H_u}$, can run freely, independently of the other soft scalar masses. In this, so-called {\bf non-universal Higgs model (NUHM)}\cite{Berezinsky:1995cj,Nath:1997qm}, the $\mu$ parameter is also less constrained, since both $\mu$ and the Higgs pseudoscalar mass $m_A$ are determined by $m_{H_d}$ and $m_{H_u}$ via the conditions of the electroweak symmetry breaking. In particular, this gives one more freedom in satisfying both the DM relic density and flavor observables\cite{Roszkowski:2009sm,Ellis:2009ai,Arbey:2011ab,Strege:2012bt,Kowalska:2013hha,Roszkowski:2014wqa,Athron:2017qdc}. 

On other hand, a simultaneous fitting of the constraints from $\deltaamu$, $m_h$ and $\Omega_\chi h^2$, favors GUT-constrained scenarios with {\bf non-universal gaugino masses (NUGM)}\cite{Kowalska:2015zja} (see also recent study\cite{Wang:2018vrr}). In this case, for large values of the soft gluino mass $M_3$, the Higgs boson mass $m_h$ can be driven to $m_h\simeq 125\gev$ via loop-induced corrections and the running of the stop mass. At the same time, the bino- or wino-like lightest neutralino can be much lighter, as required by $\deltaamu$. Specific gaugino mass unification patterns can also lead to reduced fine-tuning in supersymmetric models\cite{Kane:1998im,BasteroGil:1999gu,Abe:2007kf,Horton:2009ed} and simultaneously allow one to fit the DM relic density constraint\cite{Kowalska:2014hza}. In our scans, we analyze two such respective scenarios, in which the soft bino and wino masses are either kept equal at the GUT scale $M_1=M_2\neq M_3$ ({\bf NUGM1}), or are all varied independently ({\bf NUGM2}); cf. Ref.\cite{Kowalska:2015zja}.

Finally, in addition to treating $\mone$, $\mtwo$ and $\mthree$ as free parameters, we allow the soft squark and slepton masses to acquire non-universal values at the GUT scale. We first consider the model in which the NUHM boundary conditions are further modified by an inclusion of a small D-term contribution $m_D^2$\cite{Kawamura:1994ys,Kolda:1995iw} such that $m_Q^2 = m_U^2 = m_E^2 = m_{16}^2+M_D^2$, $m_D^2=m_L^2 = m_{16}^2-3 M_D^2$ and $m_{H_u,H_d}^2 = m_{10}^2\mp 2M_D^2$. Here, we have assumed an {\bf  {\boldmath $SO(10)$}-inspired}\cite{Georgi:1974my,Fritzsch:1974nn} GUT-constrained framework in which the universal masses in the fermionic \textbf{16} and bosonic \textbf{10} representations are denoted by $m_{16}^2$ and $m_{10}^2$, respectively. All the three gaugino masses are allowed to vary freely. The same is true for the last model that we consider which we call an $SU(5)$-inspired model\cite{Georgi:1974sy}. In this case, the Higgs sector soft masses $m_{H_u}$ and $m_{H_d}$ are set free. Other soft scalar masses are set by the common masses of the $\mathbf{\bar{5}}$ and $\textbf{10}$ representations $m_5$ and $m_{10}$, respectively, to be $m_Q^2 = m_U^2 = m_E^2 = m_{10}^2$ and $m_D^2 = m_L^2 = m_5^2$.

\def\arraystretch{1.5}
\begin{table}
\begin{tabular}{|c|c|}
\hline
Model & Parameters \\
\hline
\hline
CMSSM & $m_0$, $m_{1/2}$, $A_0$, $\tan\beta$\\
NUHM & $m_0$, $m_{H_d}^2$, $m_{H_u}^2$, $m_{1/2}$, $A_0$, $\tan\beta$\\
NUGM1 & $m_0$, $M_1=M_2$, $M_3$, $A_0$, $\tan\beta$\\
NUGM2 & $m_0$, $M_1$, $M_2$, $M_3$, $A_0$, $\tan\beta$\\
$SO(10)$-inspired & $m_{16}$, $m_{10}$, $M_D$, $M_1$, $M_2$, $M_3$, $A_0$, $\tan\beta$\\
$SU(5)$-inspired & $m_{10}$, $m_{5}$, $m_{H_d}^2$, $m_{H_u}^2$, $M_1$, $M_2$, $M_3$, $A_0$, $\tan\beta$\\
\hline
\end{tabular}
\caption{Model parameters of the GUT-constrained scenarios examined in this paper.\label{tab:parameters}}
\end{table}

\begin{table}
\begin{tabular}{|c|c|}
\hline
Parameter & Range \\
\hline
\hline
\multirow{3}{*}{Scalar masses} & $100 < m_0, m_{16}, m_{10,SU(5)}, m_5< 4000$\\
 & $-10000^2 < m_{H_d}^2, m_{H_u}^2, m_{10,SO(10)}^2 < 10000^2$\\
 & $0 < 3 M_D^2 < m_{16}^2 - (100\gev)^2$\\
\hline
\multirow{3}{*}{Gaugino masses} & $100 < m_{1/2} < 4000$\\
 & $-4000 < M_1,M_2 < 4000$\\
 & $700 < M_3 < 10000$\\
\hline
Trilinear couplings & $-10000 < A_0 < 10000$\\
\hline
Higgs vev ratio & $2 < \tan\beta < 62$ \\
\hline
\end{tabular}
\caption{The ranges of the model parameters that we scan over (all dimensionful parameters given in GeV). In our scans we use flat priors. 
\label{tab:ranges}}
\end{table}

\subsection{\label{sec:simulation}Parameter scan}

In our simulations, we use the \texttt{Multinest}\cite{Feroz:2007kg,Feroz:2008xx} package and generate points in the parameter space with flat priors. The scans are driven by the likelihood functions that contain the observables listed in Table~\ref{tab:observables}. We employ \texttt{SOFTSUSY} to obtain the supersymmetric spectrum\cite{Allanach:2001kg} and decay branching fractions\cite{Allanach:2017hcf}. The Higgs boson mass is calculated with the \texttt{Himalaya} package\cite{Harlander:2017kuc}, which includes three-loop corrections following Ref.\cite{Kant:2010tf}. The properties of the lightest SM-like Higgs boson are further constrained with the use of \texttt{Higgsbounds}\cite{Bechtle:2008jh,Bechtle:2020pkv} and \texttt{Higgssignals}\cite{Bechtle:2013xfa,Bechtle:2020uwn}. 

We use \texttt{Micromegas}\cite{Belanger:2001fz,Belanger:2004yn,Belanger:2020gnr} to calculate the relic density of the lightest neutralino DM and to include direct detection constraints from PICO-60\cite{Amole:2019fdf} and Xenon1T\cite{Aprile:2018dbl}. When comparing the DM relic density to the Planck data\cite{Aghanim:2018eyx}, we assume $10\%$ theoretical uncertainty. We study both scenarios with $\Omega_\chi h^2\simeq \Omega_{\textrm{DM}}^{\textrm{tot}}h^2$ and $\Omega_\chi h^2\lesssim \Omega_{\textrm{DM}}^{\textrm{tot}}h^2$. In the latter case, we assume that the lightest neutralino contributes only partially to the total DM relic density.\footnote{In the regions of the parameter space that correspond to sub-TeV neutralino DM with its relic density set by co-annihilations with sleptons, the Sommerfeld enhancement (SE)\cite{Sommerfeld,ArkaniHamed:2008qn} has typically an impact on $\Omega_\chi h^2$ of order up to a few percent\cite{Hryczuk:2011tq}. This is within the theoretical uncertainty that we impose on $\Omega_\chi h^2$ in our scans. A larger effect could be observed for light wino DM, but in this case, the present-day indirect bounds can be avoided thanks to suppressed relic density.} \texttt{Superiso Relic}\cite{Arbey:2018msw} is employed to test the points in the parameter space against the searches for DM-induced $\gamma$-rays in FermiLAT\cite{Fermi-LAT:2016uux} and to analyze the cosmic-ray anti-proton bounds from the AMS-02 experiment\cite{Aguilar:2016kjl}; cf. also Refs\cite{Cuoco:2017iax,Boudaud:2014qra}.

\texttt{Superiso Relic}\cite{Arbey:2018msw} is also used to calculate flavor observables. In our analysis, we include the leptonic $B_s\to\mu^+\mu^-$ and radiative $B\to X_s\gamma$ decays. In the former case, we follow Ref.\cite{Altmannshofer:2021qrr} and employ the updated world average value of the relevant branching fraction after recent LHCb measurement\cite{LHCb:Bsmumu}. For the radiative decay, we use the average value as determined by the HFLAV collaboration\cite{Amhis:2019ckw}. Instead, we do not take into account observables related to semi-leptonic $B$ meson decays (for recent discussion, see Ref.\cite{Cornella:2021sby} and references therein). In particular, these can lead to lepton flavor universality violation and the corresponding measured $R_K$ factor is now in a $3.1\sigma$ tension with the SM prediction\cite{Aaij:2021vac}. It is known, however, that all these deviations, if fully confirmed with the future data, cannot be simultaneously resolved in the framework of the MSSM\cite{Altmannshofer:2014rta}. Since our focus in this study is on $\deltaamu$, we leave an updated analysis of the semi-leptonic $B$-meson-decay anomalies for future studies related to scenarios going beyond the MSSM.

Direct LHC searches play an important role in constraining the allowed values of $\deltaamu$ in supersymmetric models. We treat them with  \texttt{CheckMATE}\cite{Drees:2013wra,Dercks:2016npn}
and employ the analyses implemented in version \texttt{2.0.23} of the package; cf. Ref.\cite{Chakraborti:2020vjp} for further discussion about relevant analyses. In addition, we test our scenarios against the search for compressed supersymmetric spectra\cite{Aad:2019qnd} by applying the cuts in the ($m_\chi,m_{\chi_1^\pm}$) and ($m_\chi,m_{\ell^\pm}$) planes for the degenerate chargino and slepton cases; cf. Ref.\cite{Chakraborti:2020vjp} for a recent discussion. If the lightest neutralino is wino-like with the mass $m_\chi\sim \textrm{a few hundred GeV}$ and a very small mass splitting, $m_{\chi^\pm_1}-m_{\chi}\sim \mathcal{O}(100~\mev)$ then further important bounds come from searches for disappearing tracks\cite{Aaboud:2017mpt,Sirunyan:2020pjd}. Such scenarios do not typically play a dominant role in determining the allowed range of $\deltaamu$ in the GUT-constrained models of our interest. However, we incorporate them in our analysis, as they could become important for the fully degenerate gaugino mass patterns and the DM relic density constraint treated as an upper limit. To this end, we rely on the approximate parametrization of the wino neutralino-chargino mass splitting at the two-loop level\cite{Ibe:2012sx} and apply the cuts following Ref.\cite{Chakraborti:2021kkr}.

\begin{table}
\begin{tabular}{|c|c|}
\hline
\multirow{2}{*}{$\ $Collider$\ $} & $m_h = (125.10\pm 0.14$\cite{Zyla:2020zbs}$\pm 2~[\textrm{th.}])\gev$\\
 & \texttt{Higgsbounds}\cite{Bechtle:2008jh,Bechtle:2020pkv}, \texttt{Higgssignal}\cite{Bechtle:2013xfa,Bechtle:2020uwn}\\
 & \texttt{CheckMATE}\cite{Drees:2013wra,Dercks:2016npn}  + Refs\cite{Aaboud:2017mpt,Aad:2019qnd,Sirunyan:2020pjd}\\
\hline
\multirow{2}{*}{$\ $DM$\ $} & $\Omega_{\textrm{DM}}^{\textrm{tot}}h^2 = 0.12\pm 0.001$\cite{Aghanim:2018eyx}$ \pm 10\%~[\textrm{th.}]$\\
 & bounds from direct \& indirect searches (see text)\\
\hline
\multirow{2}{*}{$\ $Flavor$\ $} & $\textrm{BR}(B\to X_s\gamma) = (3.32 \pm 0.15)\times 10^{-4}$\cite{Amhis:2019ckw}\\
 & $\textrm{BR}(B_s\to \mu^+\mu^-) = (2.93 \pm 0.35)\times 10^{-9}$\cite{Altmannshofer:2021qrr}\\
 \hline
 \hline
Nuisance & $m_t = 172.80\pm 0.40$\cite{TevatronElectroweakWorkingGroup:2016lid,Aaboud:2018zbu,Sirunyan:2018mlv}\\
\hline
\end{tabular}
\caption{The constraints included in the global likelihood function in our scans. The top quark mass in the bottom row is treated as a nuisance input parameter, which is drawn from a Gaussian distribution with the central value and error as indicated in the table.\label{tab:observables}}
\end{table}

Last but not least, we calculate $\deltaamu$ with the \texttt{GM2Calc}\cite{Athron:2015rva} code which employs two-loop corrections following Refs\cite{vonWeitershausen:2010zr,Fargnoli:2013zia,Bach:2015doa}. This is not included in the likelihood function that drives the numerical scans. Instead, we store the relevant values for each point in the simulations and later analyze the allowed ranges of $\deltaamu$ that are favored by other constraints listed in Table~\ref{tab:observables}.

%%%%%%%%%%%%
\subsection{\label{sec:results}Results}

In this section we present results of our numerical analysis. We begin with a general discussion about the GUT-constrained models considered in this paper. Next we focus on the models NUGM1 and the $SU(5)$-inspired model, as they represent, respectively, the cases of a minimum and a maximum number of relevant parameters (compare Table~\ref{tab:parameters}) that allow one to obtain large enough corrections to the anomalous magnetic moment of the muon.

\paragraph{GUT-constrained supersymmetry and $\deltaamu$}

\begin{figure}%[htb!]                                                       
%\vspace{2em}
  \centering
  \begin{subfigure}[b]{0.48\linewidth}
    \centering
    \includegraphics[width=1.0\textwidth]{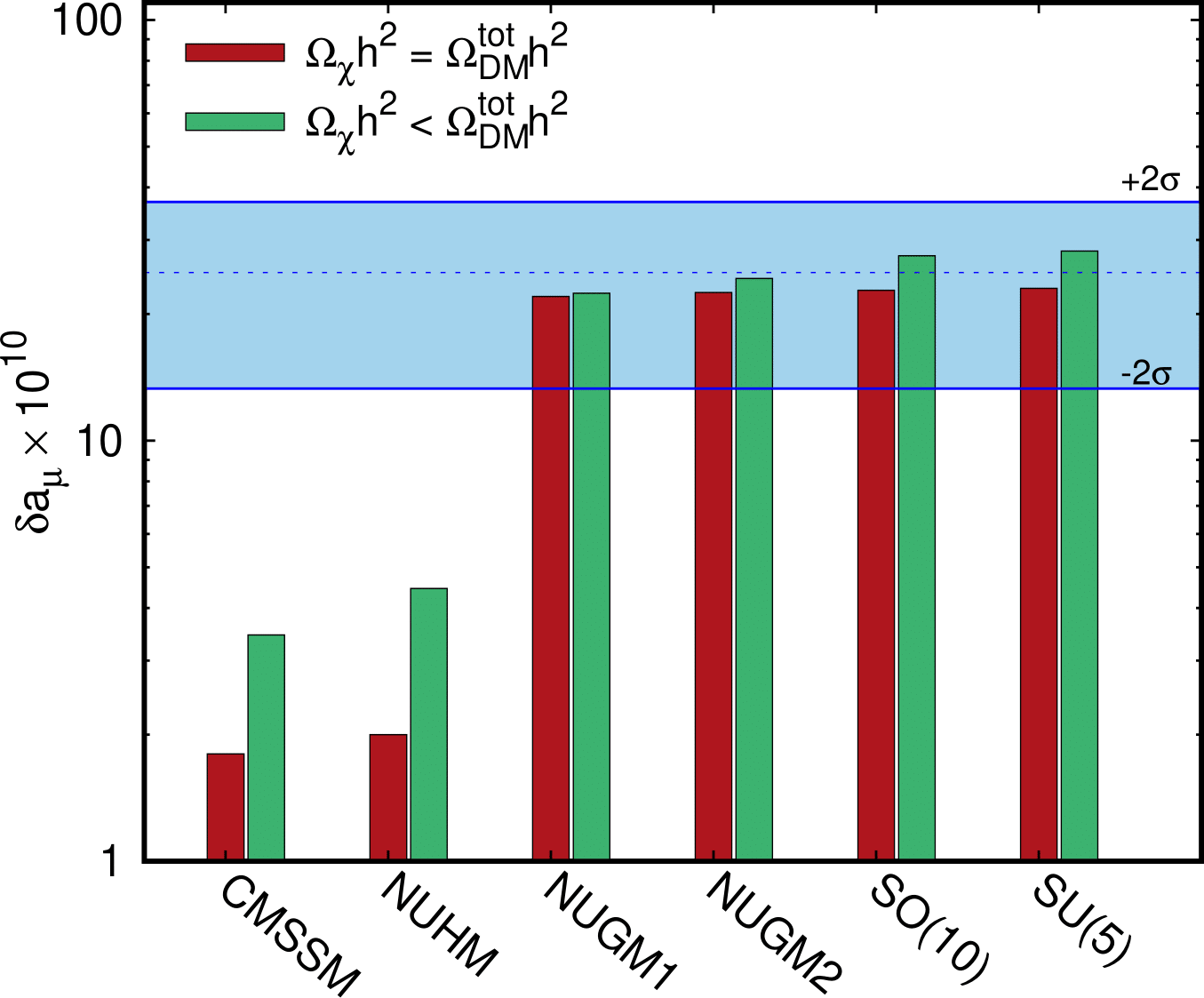}
%    \caption{}
%  \label{(a)}
  \end{subfigure}
  ~
  \begin{subfigure}[b]{0.48\linewidth}
    \centering
%lr     \includegraphics[width=1.0\textwidth]{plots/mixed_neuloop.png}
%    \caption{}
%  \label{(b)}
  \end{subfigure}
\caption{\small Histogram showing a range of values of $\deltaamu$ for scan points in the unified SUSY models examined in this work. A blue horizontal band shows the $\pm2\sigma$ range of Eq.~\ref{eq:fnalgmtwo}. In the left (red) columns the relic density constraint on the LSP from Table~\ref{tab:observables} has been  imposed fully, while in the right (green) columns only as an upper limit. 
}
\label{fig:histogramwithfulldm}
\end{figure}

In Fig.~\ref{fig:histogramwithfulldm} we present a histogram showing a range of values of $\deltaamu$ for scan points in the models examined in this paper. 
We select the points by applying a combined $2\sigma$ cut on the observational constraints included in the global likelihood function. The figure confirms the known result that in the CMSSM and the NUHM the \deltaamu\ result cannot be explained because the sleptons are too heavy. In contrast, in the two versions of the NUGM, the $SO(10)$-inspired and the $SU(5)$-inspired models some light EW-inos are allowed and can provide sufficient contribution to \deltaamu. 

%%%%%%%%%%%%%%%%%%%%%%%%%%% F I G U R E %%%%%%%%%%%%%%%%%%%%%%%%%%%%%%                                                                                  

As mentioned above, the improvement in fitting $\deltaamu$ in the latter four models is primarily a result of additional freedom in setting the gluino mass at the GUT scale independent of the masses of the wino and the bino. This allows one to satisfy the LHC bounds from direct searches for colored superpartners and from the measurements of the properties of the Higgs boson, which both favor the supersymmetric color sector to lie in the few-\tev\ mass scale. In fact, in our scans the largest values of $\deltaamu$ correspond to large gluino and squark masses, typically $m_{\tilde{g}}\gtrsim 5~\tev$ and $m_{\tilde{q}}\gtrsim 4~\tev$. On the other hand, values of $\deltaamu$ within the current $\pm 2\sigma$ bound can be obtained  also for significantly smaller gluino and squark masses, between $2$ and $3~\tev$. At the same time, the slepton and neutralino sector of the models can be kept much lighter as required by fitting the anomaly. 

Additional smaller differences between the models come from introducing more freedom in the scalar or gaugino masses that go beyond the simplest NUGM scenario; cf. Table~\ref{tab:parameters}, and by treating the relic density constraint only as an upper bound (the difference between red and green histograms in the plot). Notably, these changes do not drastically affect  the allowed ranges of $\deltaamu$, and the minimal NUGM1 model could already be sufficient to satisfy the bound given current uncertainties. On the other hand, less universal models allow one to find more fine-tuned solutions to the anomaly by employing specific mass patterns between neutralinos of different composition, charginos and sleptons.

\paragraph{The lightest neutralino composition and mass} 

\begin{figure}%[htb!]                                                       
%\vspace{2em}
  \centering
  \begin{subfigure}[b]{0.48\linewidth}
    \centering
    \includegraphics[width=1.0\textwidth]{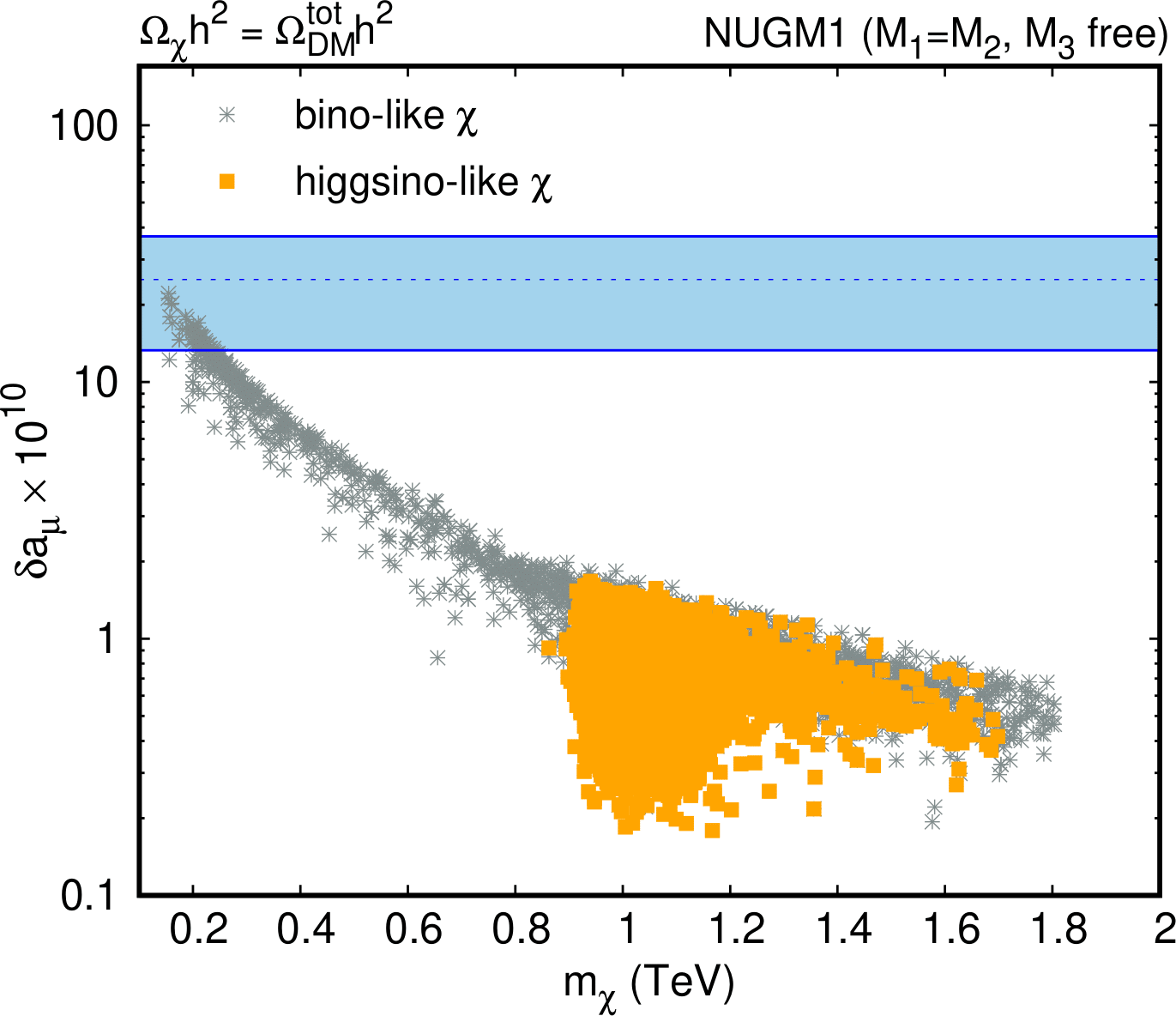}
%    \caption{}
%  \label{(a)}
  \end{subfigure}
  ~
  \begin{subfigure}[b]{0.48\linewidth}
    \centering
     \includegraphics[width=1.0\textwidth]{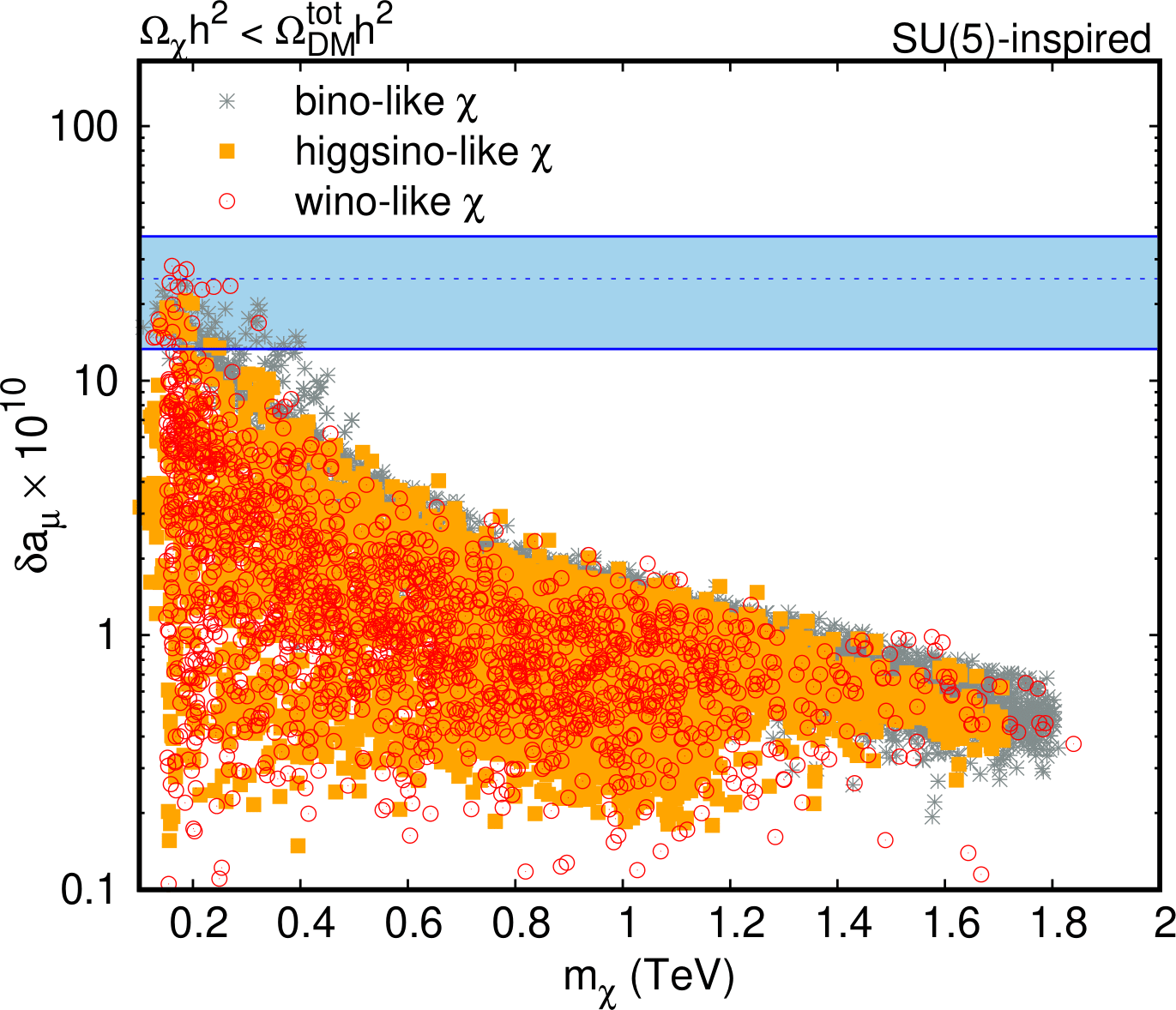}
%    \caption{}
%  \label{(b)}
  \end{subfigure}
\caption{\small 
In a $(\mchi,\deltaamu)$ plane we present a scatter plot of points from our scan that survive a global $2\sigma$ cut on the constraints given in Table~\ref{tab:observables}, showing different predominant compositions of the neutralino LSP: bino (grey stars), higgsino (beige squares) and wino (red circles) in the NUGM1 model (left window) and the $SU(5)$-inspired model (right window). The $\pm2\sigma$ range of $\deltaamu$ has also been marked.
}
\label{fig:mchideltaamu}
\end{figure}

In Fig.~\ref{fig:mchideltaamu} we display the composition of the neutralino LSP in a $(\mchi,\deltaamu)$ plane for the same selection of points as in Fig.~\ref{fig:histogramwithfulldm} but two models only: the NUGM1 with the requirement that the LSP provides all the DM in the Universe (left window) and the $SU(5)$-inspired model with the LSP being possibly a subdominant DM component (right window). In the NUGM1 model only a bino-like LSP with mass $\mchi\lesssim 300\gev$ is consistent with Eq.~\ref{eq:fnalgmtwo}. Since the higgsino is subdominant DM when its mass is less than $~1\tev$, it appears in the plot only for $\mchi \gtrsim 1\tev$ and below the $2\sigma$ bound on $\deltaamu$. The wino DM is not present in this model, since we assume $M_1=M_2$ at the GUT scale. If we relaxed the DM density constraint to an upper limit only, the $\deltaamu$ fit could be slightly improved in this model for the bino LSP, as  it is clear from Fig.~\ref{fig:histogramwithfulldm}.

%%%%%%%%%%%%%%%%%%%%%

The correct value of the DM relic density for bino DM in the NUGM1 model can be obtained because of coannihilations with light sleptons. Although typically  the lightest of them is the lighter stau, the other sleptons, including smuons and a muon sneutrino, also have similar masses, as dictated by the common scalar mass at the GUT scale. Hence, for sufficiently light binos, in this scenario the points minimizing the global likelihood, in particular due to the correct value of the DM relic density, also naturally predict an increased value of $\deltaamu$. This can be seen in the left panel of Fig.~\ref{fig:mchideltaamu}. In this plot, the upper bound on $\deltaamu$ as a function of $\mchi$ is mainly driven by the maximum contributions to the anomaly that one can obtain in the neutrinalino-slepton coannihilation channel and by the requirement to fit $\Omega_\chi h^2$.

\begin{figure}%[htb!]                                                       
%\vspace{2em}
  \centering
  \begin{subfigure}[b]{0.48\linewidth}
    \centering
    \includegraphics[width=1.0\textwidth]{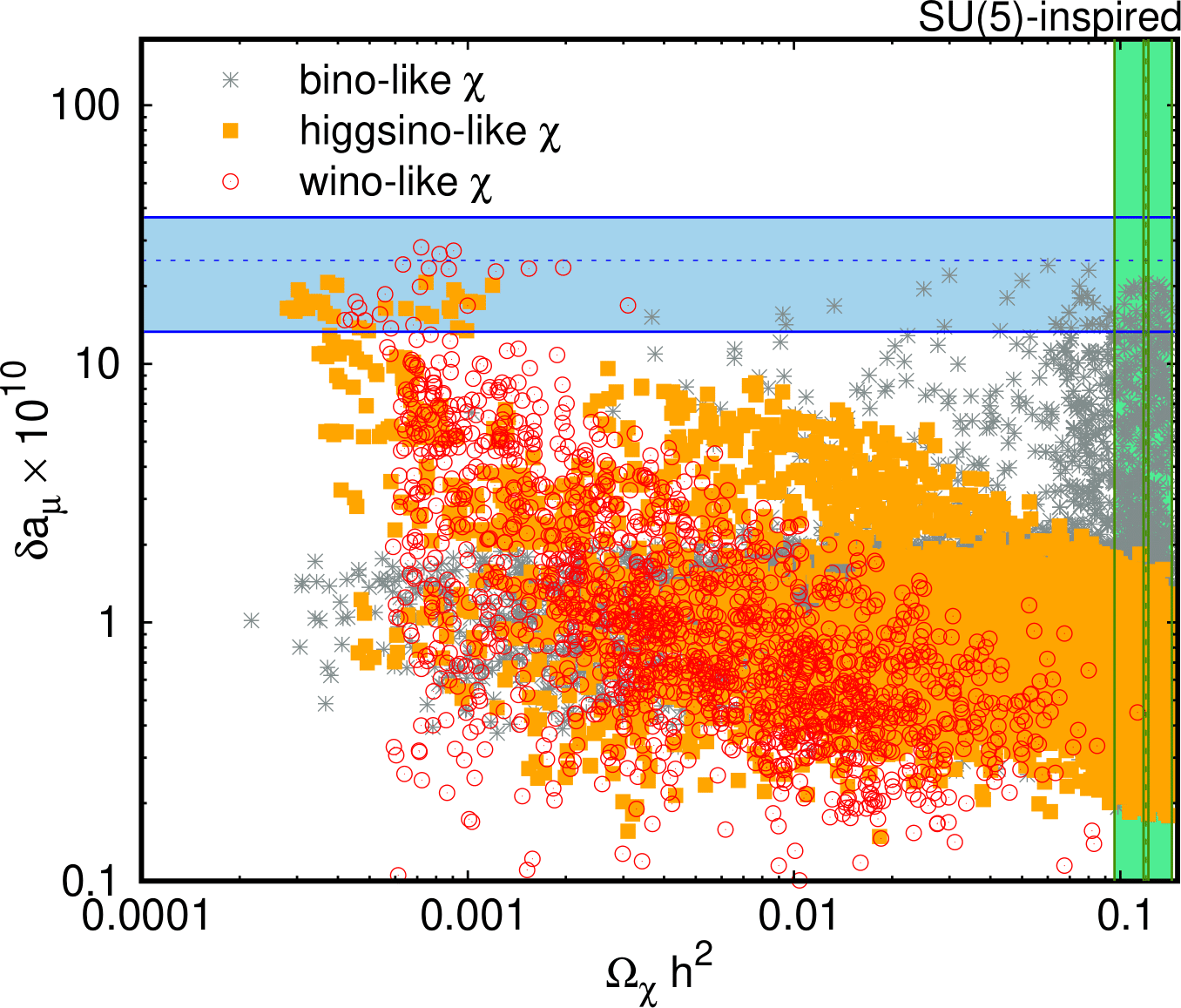}
%    \caption{}
%  \label{(a)}
  \end{subfigure}
  ~
  \begin{subfigure}[b]{0.48\linewidth}
    \centering
     \includegraphics[width=1.0\textwidth]{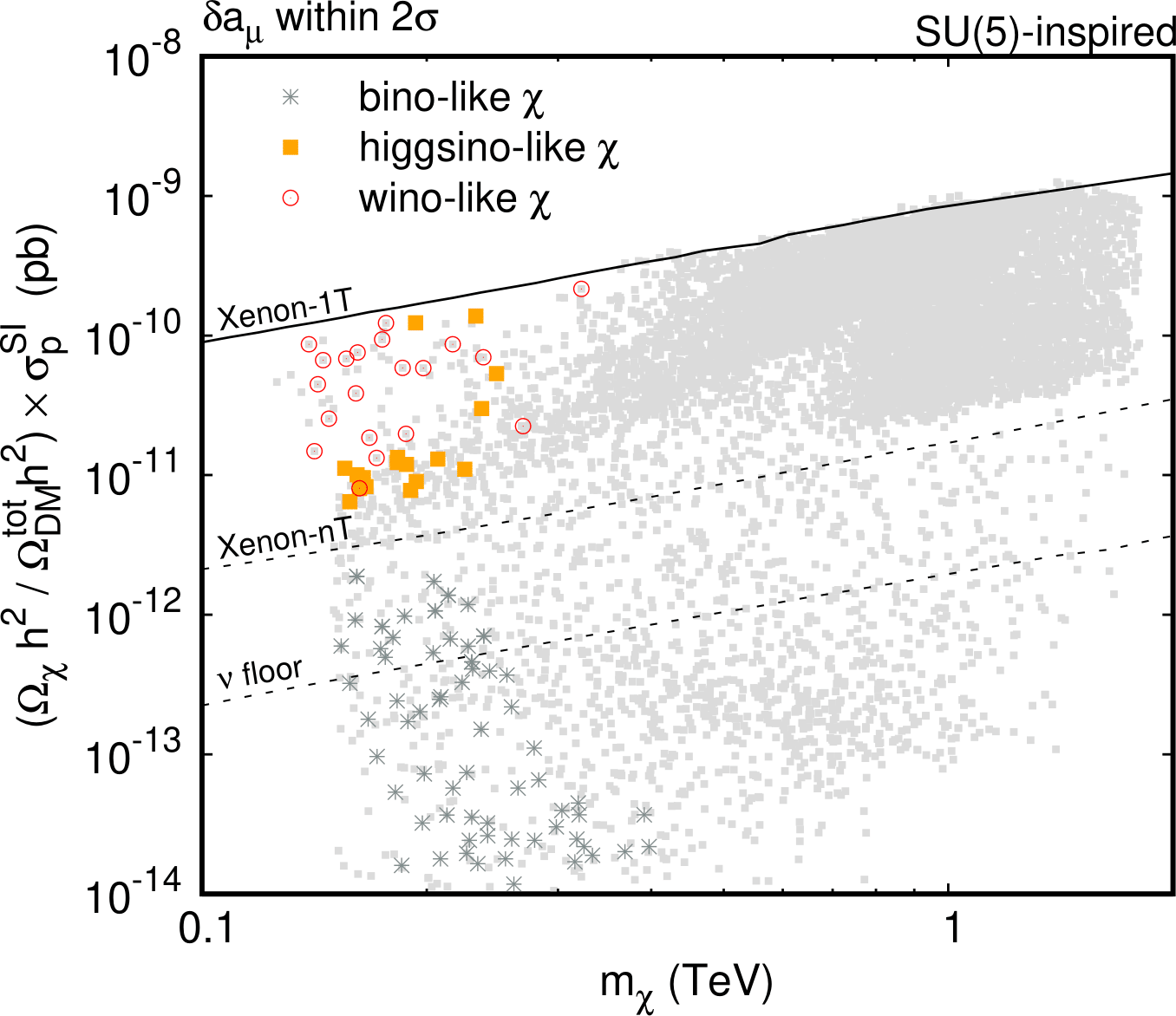}
%    \caption{}
%  \label{(b)}
  \end{subfigure}
\caption{\small 
\textsl{Left}: In a $(\abundchi,\deltaamu)$ plane we present a scatter plot of points from our scan that satisfy a global $2\sigma$ cut on the constraints given in Table~\ref{tab:observables}, showing different predominant compositions of the neutralino LSP: bino (grey stars), higgsino (beige squares) and wino (red circles) in the $SU(5)$-inspired model. The $\pm2\sigma$ ranges of both $\abundchi$ and $\deltaamu$ have also been marked. In the former case, a larger green shaded region corresponds to the assumed $10\%$ theoretical uncertainty, while a small observational uncertainty is also marked as a thick line in the center of this band. As can be seen, while for some bino points the relic density can reach the correct range, all the higgsino and wino points correspond to strongly subdominant DM. 
\textsl{Right}: In the $(m_\chi,\sigsip)$ plane we show all scan points (light grey) that satisfy the global $\pm 2\sigma$ cut on the constraints given in Table~\ref{tab:observables}. For (mostly higgsino and wino) points with subdominant relic density the spin-independent cross section has been rescaled by the true neutralino DM relic density, i.e. by the ratio $\abundchi / \Omega_{\textrm{DM}}^{\textrm{tot}}$. The darker points correspond to scenarios for which additionally $\deltaamu$ is satisfied within the $\pm 2\sigma$ band. We also show the current Xenon1T bound\cite{Aprile:2018dbl} (black solid line), future prospects for Xenon-nT\cite{Aprile:2020vtw} (black dashed) and the expected so-called ``neutrino floor'' related to atmospheric and diffuse supernova neutrinos\cite{Billard:2013qya}. 
}
\label{fig:oh2deltaamu}
\end{figure}

In the right window of Fig.~\ref{fig:mchideltaamu} the general pattern (compare Fig.~\ref{fig:mixed}) seen in the NUGM1 is found also in the $SU(5)$-inspired model but now we can additionally see some higgsino and wino LSP cases that can fit $\deltaamu$. They correspond, however, to very subdominant DM. This can be clearly seen in the left panel of Fig.~\ref{fig:oh2deltaamu} for the $SU(5)$-inspired model in which we present the results of our scans in a $(\abundchi,\deltaamu)$ plane. While a light LSP neutralino is required in order to give sufficient contribution to $\deltaamu$, only bino-like LSP can at the same time explain the whole DM in the Universe.
This is so mainly due to bino-slepton coannihilations, while in this model we also find cases of the lightest bino-like neutralino DM with important wino admixture, or bino-wino DM, that are also known to be able to satisfy the $\deltaamu$ anomaly in the general MSSM; see, e.g., Ref.\cite{Chakraborti:2021kkr} for a recent discussion.

In our analysis for the $SU(5)$-inspired model, the best fit points in the global likelihood correspond to $m_\chi\lesssim 400\gev$. We note, however, that if we 
relax the requirement to fit the global $2\sigma$ constraint used in our analysis and focus on fitting $\deltaamu$ only, then already in this GUT-constrained scenario we find good points in the parameter space with $m_\chi$ up to even $730\gev$. Notably, this is close to the upper limit on the neutralino DM mass that can fit the anomaly which can be seen in Fig.~\ref{fig:mixed}.
We note, however, that in specific scenarios predicting, e.g., a large gap between the stau and the other slepton masses, the neutralino DM mass can even exceed $1\tev$ while still contributing enough to $\deltaamu$; cf. Ref.\cite{Endo:2021zal}.

The allowed DM mass range will further shrink when the future Fermilab Muon g-2 experiment data would allow one to reduce the current uncertainty on $\deltaamu$. In particular, assuming the same central value but a reduced experimental error by a factor of two would render $m_\chi\lesssim 320\gev$ for good global fit points in our scans, in contrast to $m_\chi\lesssim 670\gev$ for points outside our global $2\sigma$ cut which explain the anomaly. Eventually, even stronger reduction of the uncertainty would have a strong impact on the allowed region of the parameter space. This would, however, depend also on the progress in reducing the theoretical uncertainty in the SM prediction of this observable.

In the discussion above, we have focused on the NUGM1 model which is 
the most constrained scenario studied in this paper that still allows one to fit $\deltaamu$, as well as on the least constrained case, the $SU(5)$-inspired model, that also has this property. It is worth mentioning, though, that a general discussion about the allowed DM scenarios relevant for the latter case remains also valid in the NUGM2 and the $SO(10)$-inspired models. In particular, already in NUGM2 we can find the low DM relic density points with predominantly wino DM. This is due to relaxing the unification condition between the bino and wino masses at the GUT scale, $M_1\neq M_2$. Smaller differences between these models, seen in Fig.~\ref{fig:histogramwithfulldm}, are caused by an increasingly easier accommodation of the global $2\sigma$ bounds in increasingly less unified scenarios. These details are, however, not crucial for our discussion.

\paragraph{Discovery prospects in DM and LHC searches} 

The observed pattern of $\chi$ mass and relic density has strong implications for neutralino DM searches. In the right panel of Fig.~\ref{fig:oh2deltaamu} we present the prospects for detecting such DM in direct detection searches. We focus on the spin-independent cross section, $\sigma_p^{\textrm{SI}}$, which is additionally rescaled by the ratio of (possibly subdominant) neutralino DM relic density to the total DM relic density, $\abundchi/\Omega_\chi^{\textrm{tot}}$. As can be seen, from all the points in the $SU(5)$-inspired model that satisfy the global $2\sigma$ constraints discussed in section~\ref{sec:simulation}, only scenarios predicting low DM mass survive the $\deltaamu$ constraint and these are marked with color. 

%%%%%%%%%%%%%%%%%%%%%

Even though such higgsino and wino points correspond to only $\mathcal{O}(1\%)$ of the total DM relic density, their DD interaction rates are still large enough to be within the reach of the forthcoming Xenon-nT detector\cite{Aprile:2020vtw}. In contrast, bino-like DM (dark-grey crosses in right panel of Fig.~\ref{fig:oh2deltaamu}) will generally remain beyond the reach of detectors like Xenon-nT. They can even give signal rates below the expected background from atmospheric and diffuse supernova neutrinos\cite{Billard:2013qya}.

On the other hand, it will be challenging to detect underabundant higgsino or wino DM in indirect detection experiments due to too low annihilation rates. The same remains typically true for bino-dominated points with low annihilation cross section, $\sigv$, for which the correct value of $\abund$ is obtained thanks to coannihilations with light sleptons. In few cases, in which the present-day annihilation cross section of bino DM is increased due to e.g. a resonance via the pseudoscalar exchange, the detection rates might become within the reach of future $\gamma$-ray searches, although this would depend on the shape of the DM profile towards the Galactic Center. See Refs\cite{Cuoco:2017iax,Hryczuk:2019nql} for further relevant discussions about such searches employing cosmic-ray and $\gamma$-ray observations. 

The nearly degenerate ($m_\chi,m_{\chi_1^\pm}$) and ($m_\chi,m_{\ell^\pm}$) spectra
in the higgsino/wino- and bino-dominated LSP cases imply that searches specifically designed
to look for compressed spectra\cite{Aad:2019qnd} at future colliders may probe these scenarios further. A projected reach for various future colliders for the higgsino case can be found in\cite{Berggren:2020tle}. Depending on the masses of the sleptons in the higgsino case, the improved slepton pair production searches\cite{Aad:2019vnb} leading to two leptons and missing energy in the final state may also become important. Similarly, for the bino case, stronger bounds on gaugino pair-production searches\cite{Aaboud:2018jiw} looking for three leptons and missing energy in the final state can prove to be crucial. Prospects
for various EW SUSY searches at the HL-LHC and HE-LHC can be found in\cite{CidVidal:2018eel}.

\section{Conclusions}\label{sec:conc}
The muon anomalous magnetic moment anomaly has been providing the only hint for a low scale of BSM physics in a few hundred GeV regime, which has been at odds with the general trend of pushing it up to, and above, the TeV scale. For this reason it has been difficult to accommodate in the simplest models of SUSY defined in terms of the smallest number of parameters, like the CMSSM or the NUHM. The recent confirmation of \deltaamu\ gives it a more solid ground and brings very strong implications for SUSY searches at the LHC and for DM. In the models discussed in this paper the SUSY spectra are quite compressed, making them difficult to access at the LHC. The \gmtwo\ anomaly can be consistent with bino-like DM being all DM. Alternatively, the same can be achieved for higgsino-like LSP but only when it is very subdominant, constituting only a few per cent of the DM in the Universe. These implications for DM are quite general and applicable to the general MSSM. In this paper we showed that these solutions can be realised in some GUT-inspired models featuring gaugino mass non-unification, but not (as has been well known) in the simplest frameworks of the CMSSM or the NUMH.

{\bf NOTE ADDED:} After our analysis had been completed, we became aware of Refs.\cite{iwamoto2021winohiggsino,wang2021gutscale} 
which partly overlap with our paper. In Ref.\cite{iwamoto2021winohiggsino}  in a seven-parameter model with non-universal gaugino masses the case of the wino-higgsino for $\mone\gg\mtwo\sim\mu$ was mostly considered which is only a very specific, and rather fine tuned, case. In Ref.\cite{wang2021gutscale} some specific non-universal gaugino mass patterns were considered but no implications for DM composition and prospects for direct detection were examined.

%%%%%%%%%%%%%%%%%%%%%%%%%%%%%%%%%%%%%%%%%%%%%%%%%%%%%%%%%%%%%%%%%%%%%%
\acknowledgments
This work is supported by the project AstroCeNT, Particle Astrophysics Science and Technology Centre, carried out within the International Research Agendas programme of the Foundation for Polish Science financed by the European Union under the European Regional Development Fund. LR is also supported in part by the National Science Centre, Poland, research grant No. 2015/18/A/ST2/00748. ST is supported in part by the Polish Ministry of Science and Higher Education through its scholarship for young and outstanding scientists (decision no 1190/E-78/STYP/14/2019). The use of the CIS computer cluster at the National Centre for Nuclear Research in Warsaw is gratefully acknowledged. 
%%%%%%%%%%%%%%%%%%%%%%%%%%%%%%%%%%%%%%%%%%%%%%%%%%%%%%%%%%%%%%%%%%%%%%
\bibliographystyle{JHEP}
\bibliography{biblio}

\end{document}